\documentclass{article}
\pdfoutput=1
\usepackage{lscape}
\usepackage{lipsum}
\usepackage{enumerate}
\usepackage{amsfonts}
\usepackage{amsmath}
\usepackage{comment}
\usepackage{textcomp}
\usepackage{amsthm, tikz}
\usepackage{amssymb}
\usepackage{color}
\usepackage{graphicx}
\usepackage{pdflscape}
\usepackage{afterpage}
\usepackage[matrix,arrow,curve]{xy}
\usepackage{array}

\makeatletter
\renewcommand*{\@cite@ofmt}{\bfseries\hbox}
\makeatother

\usepackage{braids}

\makeatletter
\renewcommand*{\@cite@ofmt}{\bfseries\hbox}
\renewcommand{\L}{\mathcal{L}}
\newcommand{\K}{\mathcal{K}}
\newcommand{\R}{\mathcal{R}}

\makeatother

\def\be{\begin{eqnarray}}
\def\ee{\end{eqnarray}}

\hoffset= - 1.2in         

\begin{document}

\title{\vspace{0.1cm}{\Large {\bf Multi-Colored Links From 3-strand Braids Carrying Arbitrary Symmetric Representations}\vspace{.2cm}}
\author{
{\bf Saswati Dhara$^{a}$},\ {\bf A. Mironov$^{b,c,d}$}, \ {\bf A. Morozov$^{b,c,d}$}, \ {\bf An. Morozov$^{c,d,e}$}, \\   {\bf P. Ramadevi$^{a}$}, \ {\bf Vivek  Kumar Singh$^{a}$},{\bf A. Sleptsov$^{c,d,f}$}
}
\date{ }
}

\maketitle

\vspace{-5.9cm}

\begin{center}
\hfill FIAN/TD-05/18\\
\hfill IITP/TH-04/18\\
\hfill ITEP/TH-06/18\\
\end{center}

\vspace{4.2cm}

\begin{center}

$^a$ {\small {\it Department of Physics, Indian Institute of Technology Bombay, Mumbai 400076, India}}\\
$^b$ {\small {\it Lebedev Physics Institute, Moscow 119991, Russia}}\\
$^c$ {\small {\it ITEP, Moscow 117218, Russia}}\\
$^d$ {\small {\it Institute for Information Transmission Problems, Moscow 127994, Russia}}\\
$^e$ {\small {\it MIPT, Dolgoprudny 141701, Russia}}\\
$^f$ {\small {\it Laboratory of Quantum Topology, Chelyabinsk State University, Chelyabinsk 454001, Russia }}
\end{center}

\vspace{1cm}

\begin{abstract}
Obtaining colored HOMFLY-PT polynomials for knots from 3-strand braid carrying arbitrary $SU(N)$ representation is still tedious.  For a class of rank $r$ symmetric representations, $[r]$-colored HOMFLY-PT $H_{[r]}$ evaluation  becomes simpler. Recently \cite{IMMMec},  it was shown that $H_{[r]}$, for such knots from 3-strand braid, can be constructed using the  quantum Racah coefficients (6j-symbols)  of $U_q(sl_2)$. In this paper, we generalise it to links whose components carry different symmetric representations. We illustrate the technique by evaluating multi-colored link polynomials $H_{[r_1],[r_2]}$ for the two-component link L7a3 whose components carry $[r_1]$ and $[r_2]$ colors.
\end{abstract}


\vspace{.5cm}



\section{Introduction}
\setcounter{footnote}{1}
For any knot colored by arbitrary $SU(N)$ representation, we can formally write colored knot invariants from Chern-Simons field theory. Equivalently, these invariants can also be
obtained from Reshetikhin-Turaev approach based on the theory of quantum groups and quantum $\mathcal{R}$-matrix. These invariants involve  braiding eigenvalues of the $\mathcal{R}$-matrix  and quantum Racah matrices  of  $U_q(sl_N)$. In order to see the explicit polynomial form of these invariants, we need the matrix elements of quantum Racah which are not known for all $SU(N)$ representations.

Recall,  $U_q(sl_2)$ Racah coefficients or Wigner 6j-symbols are very important  in many theoretical and mathematical physics. These coefficients appear as transformation matrix elements between two equivalent bases obtained from combining three angular momenta \cite{LL}.  They can also be viewed as duality matrices between two equivalent $SU(2)_k$ Wess-Zumino Witten conformal blocks \cite{Wit}-\cite{inds4}  where the quantum deformation $q$ is taken as $k+2$-th root of unity. Interestingly, there is a closed form analytic expression for the $U_q(sl_2)$ Racah coefficients \cite{KirResh} which can also be described in terms of q-hypergeometric function $_4\Phi_3$ \cite{Klimyk}.

Racah coefficients are well defined for both finite-dimensional \cite{Klimyk} and infinite-dimensional representations \cite{G,PT} of classical Lie groups as well as for quantum groups \cite{Klimyk}.
In this paper, we consider only irreducible finite-dimensional representations of quantum group $U_q(sl_N)$. Such representations are enumerated by Young diagrams and can be separated into two groups: with and without multiplicities. A representation $V$ of $U_q(sl_N)$ is called multiplicity-free if the decomposition of its tensor square $V\otimes V$ into irreducible components has no repeated summands. Multiplicity-free representations are enumerated by rectangular Young diagrams.  Unfortunately, analytic expression for  $U_q(sl_N)$ Racah matrix relating two equivalent bases involving  tensor product of  three $U_q(sl_N)$ representations is still an open problem.

For a special class of $U_q(sl_N)$ representations whose Young diagram has a single row(known as symmetric representations), which belong to  multiplicity-free class, 6j symbols are calculated for some examples \cite{IMMM3,AMMM,NRZ,MMSRacah}. Hence our aim in this paper is to determine $U_q(sl_N)$ Racah matrix elements arising in the tensor product of three arbitrary symmetric representations of $U_q(sl_N)$. Particularly, we establish equivalence between such $U_q(sl_N)$ Racah coefficients with Racah coefficients for $U_q(sl_2)$. In fact the proof for such an equivalence is based on the eigenvalue hypothesis supplemented
with the  multiplicity-free property which we will elaborate.

According to the eigenvalue hypothesis, all the Racah matrices for knots can be given in terms of normalized eigenvalues of the braiding $\mathcal{R}$-matrix. As a consequence \cite{IMMMec}, the Racah matrices for
3-strands carrying same $[r]$ symmetric representations turns out to be same as $U_q(sl_2)$ Racah matrix corresponding to the same set of eigenvalues. In Ref.\cite{Cab}, eigenvalue hypothesis is extended to
links from 3-strand braids.  Note that the generalisation to  links  from braids will now involve many possible $\mathcal{R}$ matrices with different normalized braiding eigenvalues. However, the identification of
$U_q(sl_N)$  Racah matrices appearing in the link invariant computation has not been related to $U_q(sl_2)$ Racah matrices. Hence, the main theme in this present paper is to consider three different symmetric
representations $[r_1]$, $[r_2]$ and $[r_3]$ on 3-strand braid and identify the  $U_q(sl_N)$ Racah matrices with the corresponding $U_q(sl_2)$ Racah matrices. This result will enable computation of multi-colored link invariants $H_{[r_1],[r_2],[r_3]}$ for three-component links and $H_{[r_1],[r_2]}$ for two-component links. In fact, the two-components links are even more interesting, since the three-component link from 3-strand braid is always an entangling of three unknots, while the two-component link can be more sophisticated.
We emphasise these invariants are  required to validate the integrality conjectures proposed within topological string context \cite{OV}-\cite{MMMSov}.

\bigskip

Our paper is organized as follows. In Section \ref{s.int} we define link invariants with the help of quantum $\R$-matrices. Particularly, we discuss the finite-dimensional symmetric representations of $U_q(sl_N)$ and their relations with quantum Racah coefficients. We also formulate eigenvalue conjecture for links whose components  are colored by different $U_q(sl_N)$  symmetric representations. In Section \ref{s.3br} we consider three-strand braids colored by symmetric representations. We  indicate  that the normalised eigenvalues of $\mathcal R$ matrices are same when we reduce the rank of the different symmetric representations on all the three strands  by same integer $n$.  Further we construct a proof, assuming eigenvalue hypothesis,  equating $U_q(sl_N)$ Racah matrix to $U_q(sl_2)$ Racah matrix. We work out the multi-colored HOMFLY-PT for a two-component link $L7a3$  in Section \ref{s.homfly}. Interestingly, we give a closed form expression for the multi-colored link invariant. In the concluding section, we summarize and suggest future directions towards  generalizations to higher strand braids.

\section{Link invariants from quantum groups}
\label{s.int}
We will focus in this section on obtaining multicolored link invariants from $m$-strand braids where the component knots could carry different representations. Particularly, we will follow Reshetikhin-Turaev approach based on the theory of quantum groups  and the quantum $\mathcal{R}$-matrix.

\subsection{$\mathcal{R}$-matrix and multi-colored link invariants}
First of all, let us  define quantum $\mathcal{R}$-matrices associated with the multi-colored $m$-strand braid. We associate a finite-dimensional representation $R_i$ , of the quantized universal enveloping algebra $U_q(sl_N)$, with $i$-th strand where we assume the quantum deformation parameter $q$ to be root of unity.  In fact, all finite-dimensional representations are representations of highest weights which can be enumerated using Young diagrams. Hence from now on we will identify these representations using the Young diagrams and follow the conventional notation. For example, the notation $[l,m,n,\ldots]$ denotes Young diagram with  $l$-boxes in the first row, $m$-boxes in the second row and so on.

\bigskip

$\bullet$ For the $U_q(sl_N)$ algebra defined using generators $\{H_i,E_i,F_i\}$, there exists a universal
$\check{\mathcal{R}}$-matrix:
\begin{equation}
\check{\mathcal{R}} = q^{\sum\limits_{i,j}C_{ij}^{-1}H_i\otimes H_j}
\prod_{\textrm{positive root }\beta} \exp_q
[( 1-q^{-1}) E_\beta\otimes F_\beta]\,.
\end{equation}
where $( C_{ij})$ is the Cartan matrix.

$\bullet$ The action of quantum ${\mathcal R}_i$ on the $U_q(sl_N)$ modules $V_i$ and
$V_{i+1}$ involves the above universal $\check{\mathcal{R}}$-matrix as well as a permutation operator as
shown below:
\begin{equation}
\label{Rmat}
\mathcal{R}_i = 1_{V_1}\otimes1_{V_2}\otimes\ldots\otimes P \check{\mathcal{R}}_{i,i+1} \otimes\ldots\otimes1_{V_m} \ \in \text{End}(V_1\otimes\ldots,\otimes V_m)~,
\end{equation}
where the permutation operation is $P(x\otimes y) = y\otimes x$  and $\check{\mathcal{R}}$ acts only on   $V_i$ and $V_{i+1}$ and the identity operation on the rest of the modules $V_{j\neq i,i+1}$.
It  is well known \cite{KirResh}, \cite{RJ}-\cite{LiuPeng} that $\mathcal{R}_1,\ldots,\mathcal{R}_{m-1}$ define a representation of the Artin's braid group $B_m$ on $m$ strands:
\begin{equation}
\begin{array}{rcl}
\pi: B_m  &\rightarrow&  \text{End}(V_1\otimes\ldots,\otimes V_m) \\
\pi(\sigma_i) &=& \mathcal{R}_i,
\end{array}
\end{equation}
where $\sigma_1,\ldots,\sigma_{m-1}$ are generators of the braid group $B_m$.
Graphically we can represent $\mathcal{R}_i$ as follows:

\begin{picture}(850,150)(-250,-105)
\put(-70,20){\line(1,0){90}}.
\put(-35,0){\mbox{$\ldots$}}
\put(-70,-20){\line(1,0){35}}
\put(-70,-40){\line(1,0){35}}
\put(-15,-20){\line(1,0){35}}
\put(-15,-40){\line(1,0){35}}
\put(-35,-60){\mbox{$\ldots$}}
\put(-70,-80){\line(1,0){90}}
\put(-35,-20){\line(1,-1){20}}
\multiput(-15,-20)(-10,-10){2}{\line(-1,-1){10}}
\put(-130,-33){\mbox{$\mathcal{R}_i$ \ \ = \ \ }}
\put(-90,17){\mbox{$V_1$}}
\put(-90,-23){\mbox{$V_i$}}
\put(-90,-43){\mbox{$V_{i{+}1}$}}
\put(-90,-83){\mbox{$V_m$}}
\end{picture}

Clearly, inverse crossing is given by  $\mathcal{R}^{{-}1}_i$. Operators $\mathcal{R}_i$ satisfy  relations of the braid group $B_m$:
\begin{equation}
\begin{array}{lll}
\textbf{commutativity property} & \mathcal{R}_i\mathcal{R}_j = \mathcal{R}_j\mathcal{R}_i, & \text{for} \ |i-j|\neq 1 \\
\textbf{braiding property} & \mathcal{R}_i\mathcal{R}_{i+1}\mathcal{R}_{i} = \mathcal{R}_{i+1}\mathcal{R}_i\mathcal{R}_{i+1}, & \text{for} \ i=1,\ldots,m-2
\end{array}
\end{equation}
Graphically, the braiding relation is equivalent to the third Reidemeister move whereas  algebraically, it is a well-known \textit{quantum Yang-Baxter equation} on quantum $\mathcal{R}$-matrix.

$\bullet$ According to Alexander's theorem,  any  link  in $\mathbb{R}^3$ can be constructed from closure of $m$-strand braid. Let $\L$ be an oriented link with $L$ components $\K_1,\ldots,\K_L$ colored by representations $[r_1],\ldots,[r_L]$ and  $\beta_{\L} \in B_m$ is a some braid whose closure gives $\L$. Then according to Reshetikhin-Turaev approach \cite{RT1}, \cite{RT2} the quantum group invariant, which is also known as colored HOMFLY polynomial, of the link $\L$ is defined as follows\footnote{The usual framing factor in front of quantum trace, which provide the invariance under the first Reidemeister move, we incorporate in $\R$-matrix by modifying its eigenvalues (\ref{evR1}).}:
\begin{equation}
\label{HMF1}
H_{[r_1],\dots,[r_L]}^{\L} =  {}
_q\text{tr}_{V_1\otimes\dots\otimes V_m}\left( \, \pi(\beta_{\L}) \, \right),
\end{equation}
where $_q\text{tr}$ is a quantum trace whose definition in the theory of quantum groups \cite{Klimyk}
for every $z \in \text{End}(V)$ is a follows:
\begin{equation}
\label{qtr}
_q\text{tr}_V(z) = \text{tr}_V(zK_{2\rho}),
\end{equation}
where $\vec{\rho}$ is the Weyl vector (half the sum of positive roots) whose relation  in terms of simple roots $\vec{\alpha}_i$ is $2 \vec{\rho}=\sum_i n_i \vec{\alpha}_i$. The explicit form of $K_{2 \rho}$ is
$$K_{2\rho} = K_1^{n_1} \, K_2^{n_2} \ldots K_{N-1}^{n_{N-1}}$$
where  $K_i = q^{\vec{\alpha}_i . {\mathbf H}}$ involving Cartan generators $H_1, H_2, \ldots
H_{N-1}$.

\bigskip

$\bullet$ Technically, more convenient is to use a modified version of the Reshtikhin-Turaev approach \cite{MMM1}-\cite{Anopath}. Let us expand tensor product of symmetric representations( denoted as Young diagram notation) $[r_1]\otimes [r_2]\otimes \ldots \otimes [r_m]$ into a direct sum of irreducible representations as shown
\begin{equation}
\label{irrdec}
\bigotimes_i [r_i] = \bigoplus_{\nu} ({\rm dim} \mathcal{M}_{Q_\nu} ^{1,2 \ldots m}) ~ Q_{\nu},
\end{equation}
where $Q_{\nu}$ denote the irreducible representations. If any irreducbible representation occurs more than once (called multiplicity),  then $\mathcal{M}_{Q_\nu}^{1,2, \ldots m}$ indicates  the subspace of the highest weight vectors sharing same highest weights\footnote{
	Recall, that if $Q_{\nu}$ is a Young diagram $Q_\nu=\{Q_{\nu_1}\geq Q_{\nu_2}\geq\ldots,Q_{\nu_l}>0\}$, then the highest weights $\vec{\omega}_{Q_{\nu}}$ of the corresponding representation are $\omega_i= Q_{\nu_i}-Q_{\nu_{i+1}} \ \forall \, i=1,\dots,l$, and vice versa $Q_{\nu_i} = \sum_{k=i}^l \, \omega_k$.}
corresponding to Young diagram $Q_{\nu} \vdash \sum_i|r_i|$. The dimension of the space $\mathcal{M}_{Q_\nu}^{1,2,\ldots m}$ is called \textit{multiplicity} of representation $Q_{\nu}$
which we indicate as $Q_{\nu,s}$ where $s$ takes values $1,2,\ldots {\rm dim} \mathcal{M}_{Q_{\nu}}^{1,2,\ldots m}$.

To evaluate quantum trace(\ref{qtr}), we need to write the states in weight space incorporating the mulitplicity as well.  We have many equivalent ways of denoting the state corresponding to the irreducible representations $Q_{\nu}$ (\ref{irrdec}).  One such state in the weight space denoted by the
following diagram

\hskip6cm
\begin{picture}(400,80)(-40,-32)
\put(0,0){\line(-1,1){30}}
\put(-24,24){\vector(1,-1){0}}
\put(0,0){\line(1,1){30}}
\put(24,24){\vector(-1,-1){1}}
\put(-10,10){\line(1,1){20}}
\put(04,24){\vector(-1,-1){1}}
\put(-20,20){\line(1,1){10}}
\put(-10,30){\vector(-1,-1){6}}
\put(-37,35){\mbox{$[r_1]$}}
\put(-13,35){\mbox{$[r_2]$}}
\put(7,35){\mbox{$[r_3]$}}
\put(27,35){\mbox{$[r_4]$}}
\put(-27,9){\mbox{$X_{\alpha}$}}
\put(-22,-1){\mbox{${Q_1}_{\beta}$}}
\multiput(2,-2)(3,-3){3}{\circle*{2}}
\put(-9,-15){\mbox{${Q_3}_{\gamma}$}}
\put(10,-10){\line(1,1){40}}
\put(44,24){\vector(-1,-1){0}}
\put(47,35){\mbox{$[r_m]$}}
\put(10,-10){\line(1,-1){10}}
\put(10,-10){\vector(1,-1){8}}
\put(17,-30){\mbox{$Q_{\nu}$}}
\end{picture}
 is written algebraically as

\begin{equation}
\label{decomp}
\vert \left(\ldots \left(( [r_1] \otimes [r_2])_{X_{\alpha}} \otimes [r_3]\right)_{Q_{1_{\beta}}} \ldots [r_m]\right)_{Q_{\nu}}\rangle^{(s)} \equiv | Q_{\nu}; Q_{\nu,s}, X_{\alpha}\rangle
\cong~| Q_{\nu,s},X_{\alpha}
\rangle \otimes |Q_{\nu}\rangle~,
\end{equation}
where $[r_1]\otimes [r_2]=\oplus_{\alpha=0}^{|r_1-r_2|} X_{\alpha}\equiv [r_1+r_2-\alpha,\alpha]$ and $Q_{\nu,s} \in \mathcal{M}_{Q_\nu}^{1,2,\ldots m}$ with the index $s$ to keep track of the different highest weight vectors sharing the same highest weight $\vec {\omega}_{Q_\nu}$.

Incidentally, the choice of state (\ref{decomp}) is an eigenstate of quantum
$\mathcal{ R}_1$ matrix:
\begin{eqnarray}
\mathcal{R}_1\vert \left(\ldots \left(( [r_1] \otimes [r_2])_{X_{\alpha}}
\otimes [r_3 ]\right)_{Q_{1_{\beta}}} \ldots [r_m]\right)_{Q_{\nu}}\rangle^{(s)}=
\lambda_{X_{\alpha},s}([r_1],[r_2])\vert Q_{\nu,s}, X_\alpha\rangle ~|Q_{\nu}\rangle\nonumber\\
~~~~=(\lambda_{{X_{\alpha}},s}([r_1],[r_2])\vert \left(\ldots \left(( [r_2] \otimes [r_1])_{X_\alpha} \otimes [r_3]\right)_{Q_{1_{\beta}}} \ldots R_m\right)_{Q_{\nu}}\rangle^{(s)}~.\label{eveqn}
\end{eqnarray}
Hence we will denote the $\mathcal R_1$ matrix which is diagonal in the above basis  as  $ \Lambda_{\mathcal  R_1}$ involving diagonal matrix elements $\lambda_{X_{\alpha},s}([r_1],[r_2])$.
These elements are the braiding eigenvalues
whose  explicit form is \cite{Klimyk,GZ}
 \begin{equation}
 \label{evR1}
       \lambda_{X_{\alpha},s} ([r_1],[r_2])=
        \begin{cases}
            \epsilon_{X_{\alpha},s} q^{\varkappa(X_{\alpha})-\varkappa({[r_1})-\varkappa({[r_2]})} & \text{if $[r_1]\neq [r_2]$ } \\
            \epsilon_{X_{\alpha},s} q^{\varkappa(X_{\alpha})-4\varkappa({[r_1]})-|r_1|N} & \text{if $[r_1]= [r_2]$}
            \end{cases}
    \end{equation}
For a representation $X_{\alpha}$ whose Young diagram is denoted by $\alpha_1\geq \alpha_2\ldots,\geq\alpha_{N-1}$, $ \varkappa_{X_{\alpha}} = \tfrac{1}{2} \sum_j \alpha_j (\alpha_j+1-2j)$
and $\epsilon_{X_{\alpha},s}$ will be $\pm 1$ if  $X_{\alpha}$ is connected to the multiplicity subspace state $Q_{\nu,s}$  and zero otherwise.
We will scale $ \lambda_{X_{\alpha},s}([r_1],[r_2])\rightarrow {\rm const}~\tilde \lambda_{X_{\alpha},s}([r_1],[r_2])$
such that
\begin{equation}
\prod_s  \tilde{\lambda}_{X_{\alpha},s}([r_1],[r_2]) = 1
\end{equation}
and use these normalised values $\tilde{\lambda}_{X_{\alpha},s}([r_1],[r_2])$ in the definition of
$\Lambda_{\mathcal R_1}([r_1],[r_2])$ in the rest of the paper.

As these quantum $\mathcal{R}$-matrices commute with any element from $U_q(sl_N)$, these  $\mathcal{R}$-matrices gets a block structure corresponding to the decomposition on irreducible components (\ref{irrdec}, \ref{decomp}), that is., it does not mix vectors from different representations. Futhermore, $\mathcal{R}$-matrix acts on $|Q_{\nu}\rangle$ as an identity operator and nontrivially on the subspace $\mathcal{M}_{\nu}^{1,2,\ldots m}$. The element $K_{2\rho}$ acts diagonally on $|Q_{\nu}\rangle$ but  as identity operator on subspace $\mathcal{M}_{\nu}^{1,2 \ldots m}$  as this space represent all possible highest weight vectors $Q_{\nu,s}$ with the same weight $\vec {\omega}_{Q_\nu}$.    Incorporating these facts  and the decomposition of states (\ref{decomp}), the colored HOMFLY (\ref{HMF1}) will become
\begin{eqnarray}
H_{[r_1],\dots,[r_L]}^{\L} \left(q,\,A=q^N \right)
&=&
\text{tr}_{V_1\otimes\dots\otimes V_m}\left( \, \pi(\beta_{\L}) \, K_{2\rho} \, \right) \ = \  \sum_{\nu} \text{tr}_{\mathcal{M}_\nu^{1,2 \ldots m}}\left( \, \pi(\beta_{\L}) \, \right) \cdot \text{tr}_{Q_{\nu}}\left( \, K_{2\rho} \, \right) \nonumber\\
&=&  \sum_{\nu} \text{tr}_{\mathcal{M}_\nu^{1,2 \ldots m}}\left( \, \pi(\beta_{\L}) \, \right) \cdot \text{qdim}{Q_\nu}~=~\sum_{\nu,s}\langle Q_{\nu,s},X_{\alpha}\vert
\pi(\beta_{\L} \vert Q_{\nu,s},X_\alpha\rangle~\text{qdim}{Q_\nu},
\label{linv}
\end{eqnarray}
where $\text{qdim}{Q_\nu}$ is a quantum dimension of the representation $Q_{\nu}$ explicitly given in terms of Schur polynomials: \begin{equation}
\text{qdim}{Q_\nu} = s_{Q_\nu}\left( x_1,\ldots,x_N \right)\Big|_{x_i=q^{N+1-2i}} = s_{Q_\nu}^{}(p_1,\dots,p_N)\Big|_{p_k=p_k^{*}} \equiv s_{Q_\nu}^{*}(A,q),
\end{equation}
where $p_k=\sum\limits_{i=1}^{N} x_i^k$ and $p_k^* = \frac{A^k-A^{-k}}{q^k-q^{-k}}$. For $s_{Q_\nu}^{*}(A,q)$ there exists a very simple and straightforward hook formula:
\begin{equation}
s_{Q_\nu}^{*}(A,q) =  \prod_{(i,j)\in R} \frac{Aq^{j-i} - (Aq^{j-i})^{-1}} {q^{\kappa(i,j)+\lambda(i,j)+1} - (q^{\kappa(i,j)+\lambda(i,j)+1})^{-1}},
\end{equation}
where $\kappa(i,j) = R_i-j-1$ and $\lambda(i,j)=R_j'-i-1$. Here $R_i$ denotes the number of boxes in the $i$-th row of the Young diagram of $Q_{\nu}$ and
$R_j'$ is the number of boxes of the $j$-th row of transpose of the Young diagram of $Q_{\nu}$.
Using eqns.(\ref{eveqn}, \ref{evR1}), we can obtain multi-colored HOMFLY-PT $H_{[r_1],[r_2]}^{\L}(q,A=q^N)$ for braid words $\mathcal R_1^{2 n}$ belonging to 2-strand braid. Note that $X_{\alpha}$ will be $Q_{\nu}$ which are multiplicity free. Hence
\begin{eqnarray}
H_{[r_1],[r_2]}^{\L}(q,A=q^N)&=&\sum_{\nu,s} \tilde{\lambda}_{X_{\alpha},s}([r_1],[r_2])^{2n} \delta_{X_{\alpha}, Q_{\nu}} \delta_{s,1} ~{\rm qdim} Q_{\nu}\nonumber\\
~&=&\sum_{\alpha} \tilde{\lambda}_{X_{\alpha},s}([r_1],[r_2])^{2n} ~ {\rm qdim} Q_{\nu}
\end{eqnarray}
 However to go beyond 2-strand braid, we need to deal with elements of
 $\mathcal R_{i\neq 1}$ as well. Hence,
to work out the trace of any braid word in the multiplicity subspaces, we need to perform transformation of the states (\ref{decomp}), using  quantum Racah coefficients,  so that  quantum $\mathcal R_{i\neq 1}$ matrices are diagonal in the transformed states. In the following subsection, we will focus on 3-strand braid and indicate the transformation between two possible states through $U_q(sl_N)$ Racah matrices and obtain the matrix form for
$\mathcal R_2$.

\subsection{Quantum Racah matrix for 3-strand braid }
Consider highest weight states of the three finite-dimensional irreducible representations $[r_1],[r_2], [r_{3}]$  of $U_q(sl_N)$. As the tensor product of these representations is associative, one has a natural isomorphism between two equivalent bases states. Hence we can unitarily relate the
two equivalent basis states as follows \cite{KirResh,Klimyk,GuJ}:
\begin{equation}
\label{asis}
|\left(\left( [r_1]\otimes [r_{2} ]\right)_{X_{\alpha}}\otimes [r_{3}]\right)_{Q_{\nu}} \rangle^{(s_1)} \xrightarrow[\text{}]{\text{U}} \vert \left([r_1] \otimes \left( [r_{2}]\otimes [r_{3}] \right)_{Y_{\beta}} \right)_{Q_{\nu}}\rangle^{(s_2)}~,
\end{equation}
where the elements of the transformation matrix $U$ (known as  quantum Racah coefficients) is
\begin{equation}
U\left[\begin{array}{cc|c}~ [r_1] & [r_2]& X_{\alpha} \\
~~ [r_3]& \bar{Q}_{\nu} &Y_{\beta}
\end{array}\right].
\end{equation}
Remember that $X_{\alpha} \otimes [r_3] \in  Q_{\nu}$ and $[r_1] \otimes Y_{\beta}\in Q_{\nu}$ and
the notation of quantum Racah coefficient means that
the representation $X_{\alpha} \in [r_1]\otimes [r_2] \cap  \bar { [r_3]}\otimes {Q}_{\nu}$ and
the representation $Y_{\beta} \in [r_2]\otimes [r_3] \cap  \bar{ [r_2]}\otimes {Q}_{\nu}$. The corresponding Racah matrix $U$ is denoted as
\begin{equation}
U\left[\begin{array}{cc}~ [r_1]&[r_2] \\
~[r_3]& \bar{Q}_{\nu} \end{array}\right]~.\label{racah3}
\end{equation}
Pictorially the above Racah matrix is drawn with three ingoing external lines $[r_i]$'s and an outgoing external line $Q_{\nu}$ (implying its conjugate representation):

\begin{picture}(250,120)(-100,-60)
\put(0,0){\line(1,0){50}}
\put(0,0){\line(-1,1){30}}
\put(-22,22){\vector(1,-1){2}}
\put(0,0){\line(-1,-1){30}}
\put(-22,-22){\vector(1,1){2}}
\put(50,0){\line(1,1){30}}
\put(72,22){\vector(-1,-1){2}}
\put(50,0){\line(1,-1){30}}
\put(50,0){\vector(1,-1){24}}
\put(-45,-30){\mbox{$[r_1]$}}
\put(-45,30){\mbox{$[r_2]$}}
\put(85,-30){\mbox{$Q_{\nu}$}}
\put(85,30){\mbox{$[r_3]$}}
\put(18,4){\mbox{$X_{\alpha}$}}
\put(130,0){\vector(1,0){40}}
\put(148,5){\mbox{$U$}}
\put(250,0){
	\put(0,-20){\line(0,1){40}}
	\put(0,-20){\line(-1,-1){30}}
	\put(-14,-34){\vector(1,1){1}}
	\put(0,-20){\line(1,-1){30}}
	\put(0,-20){\vector(1,-1){18}}
	\put(0,20){\line(1,1){30}}
	\put(14,34){\vector(-1,-1){1}}
	\put(0,20){\line(-1,1){30}}
	\put(-14,34){\vector(1,-1){1}}
	\put(-45,-40){\mbox{$[r_1]$}}
	\put(-45,40){\mbox{$[r_2]$}}
	\put(35,40){\mbox{$[r_3]$}}
	\put(35,-40){\mbox{$Q_{\nu}$}}
	\put(5,-4){\mbox{$Y_{\beta}$}}
}
\end{picture}

\noindent
Similar to the eqn.(\ref{eveqn}),  the basis state of $\mathcal {R}_2$ obeys
\begin{equation}
\label{eveqn1}
\mathcal{R}_2\vert \left([r_1]\otimes ([r_2]\otimes [r_3])_{Y_{\beta}}\right)
\rangle^{s_2}=\lambda_{Y_{\beta},s_1}([r_2],[r_3])\vert \left([r_1]\otimes ([r_3]\otimes [r_2])_{Y_{\beta}}\right)
\rangle^{s_2}~,
\end{equation}
whose diagonal matrix form  will be $ \Lambda_{\mathcal R_2}$.
The matrix form in the basis (\ref{decomp}) can be deduced using eqns.(\ref{eveqn1},\ref{asis}) as
\begin{equation}
\label{R2}
\mathcal{R}_2 = U^{\dagger}\left[\begin{array}{cc} [r_1]&[r_3] \\
~[r_2] & \bar{Q_\nu} \end{array}\right] \cdot \left( \Lambda_{\mathcal{R}_2}([r_2],[r_3]) \right) \cdot U\left[\begin{array}{cc} [r_1]&[r_2] \\ ~[r_3]& \bar{Q_\nu}
 \end{array}\right].
\end{equation}
Using the systematic approach of highest weight method \cite{MMM1}, \cite{MMMS21}-\cite{China}, the quantum Racah coefficients have been explicitly calculated for some $[r_i]$'s. However, our aim in this paper is to obtain the
Racah matrix when the symmetric representations on the 3-strands are
arbitrary $[r_i]$'s. We shall present in the next section, the proofs and details relating these  $U$ matrices to an equivalent $U_q(sl_2)$ Racah matrices.
Once the $U$ matrix elements are known, the trace in the multiplicity subspace for any braid word $\prod_i \mathcal R_1^{a_i}\mathcal R_2^{b_i}$
belonging to three-strand braid can be determined.

The methodology of formally writing the trace in multiplicity subspace presented for braid words belonging to 2-strand and 3-strand braid can be
generalised to higher strand braids.   The matrix form of quantum $\mathcal R_3, \mathcal R_4,\ldots$
in the basis(\ref{decomp}) will involve many Racah matrices. Since our focus in this paper is to compute the multi-colored link invariants from 3-strand braids carrying different symmetric representations, we  leave $m$-strand quantum $\mathcal R_{i\geq 3}$ matrices and the corresponding computation of the Racah matrices for a  future publication.

 Before we proceed to prove a neat correspondence of the quantum Racah matrix (\ref{racah3}) to $U_q(sl_2)$ Racah matrix, we will highlight the eigenvalue hypothesis implications for strands carrying different symmetric representations.

\subsection{Eigenvalue hypothesis}
\label{evcj}
In \cite{IMMMec}, the eigenvalue hypothesis states that  quantum Racah coefficients can be written in terms of the eigenvalues of $\mathcal{R}$-matrices. In fact for knots,  the Racah matrices of sizes up to $6 \times 6$ were explicitly written using the set of normalized eigenvalues of the corresponding $\mathcal{R}$-matrix \cite{IMMMec}, \cite{Univ} confirming the eigenvalue hypothesis (see also \cite{MMAl}).

For describing multi-colored links from 3-strand braid, carrying three different representations ($[r_1],[r_2],[r_3]$), there are three different quantum $\mathcal R$  eigenvalues
$\tilde{\lambda}_{X_{\alpha},s}([r_i,[r_j])$ where $i \le j$. Hence we conjecture a generalisation of the eigenvalue hypothesis applicable for links.

\textbf{Conjecture:} \textit{The Racah matrix
$U\left[\begin{array}{cc}~ [r_1] &[ r_2] \\~ [r_3] & {\bar {Q}_{\nu}} \end{array}\right]$
is expressed through 3 sets of the normalized eigenvalues  of the corresponding
 three possible $\mathcal{R}$ matrices whose diagonal form will be $\Lambda_{\mathcal R}([r_i],[r_j])$ where $i \le j=1,2,3$.}

\subsection{Signs of the eigenvalues}
\label{signev}

Formula (\ref{evR1}) defines the eigenvalues of the $\mathcal{R}$-matrices up to a sign $\epsilon_{X_{\alpha},s}$. These signs also exist in the classical limit $q=1$ when $\mathcal{R}$-matrix is just a permutation operator. $\mathcal{R}$-matrix is an operator acting from the space $[r_1]\otimes [r_2]$ to the space $[r_2] \otimes [r_1]$:
\begin{equation}
\mathcal{R}: [r_1]\otimes [r_2]\rightarrow [r_2]\otimes [r_1].
\end{equation}
If we study knots then $[r_1]=[r_2]=[r]$ and both spaces are the same ones. Here signs of the eigenvalues depend on whether highest weight vectors of the representations are \textit{symmetric} or \textit{antisymmetric} under permutation of two representations $[r_1]$ and $[r_2]$. For symmetric representation $[r]$, we can place sign as $(-1)^{\alpha}$ for irreducible representation
 $X_{\alpha}=[2r-\alpha,\alpha]$. Equivalently,  we associate $+$ sign for that $X_{\alpha}$ whose eigenvalue has highest power of $q$ and alternating signs for the other $X_{\alpha}$ in descending powers of $q$. For example, we take $R=[2]$.There are three eigenvalues, namely $\epsilon_{[4]}q^{4}$, $\epsilon_{[3,1]}q^{0}$ and $\epsilon_{[2,2]}q^{-2}$. Then the signs are $\epsilon_{[4]}=+1$, $\epsilon_{[3,1]}=-1$ and $\epsilon_{[2,2]}=+1$.

It appears that the same sign convention is applicable for eigenvalues of  $\mathcal{R}$-matrices when
$[r_1] \neq [r_2]$. In fact, this sign convention determines the signs of the Racah coefficients obtained from highest weight method. These are the signs we will use for eigenvalues of $\mathcal R$ applicable for
multi-colored links obtained from 3-strand braid.

\section{Three-strand braid colored by symmetric representations}
\label{s.3br}
In this section we consider 3-strand braids colored by arbitrary symmetric representations:$[r_1],[r_2],[r_3].$
As mentioned earlier, there will be three possible $\mathcal R$ matrices with three sets of normalised eigenvalues (\ref {evR1}) constituting the diagonal matrices ${\rm diag}\{\Lambda_{\mathcal R}([r_1],[r_2])\}$,
 ${\rm diag}\{\Lambda_{\mathcal R}([r_1],[r_3])\}$, ${\rm diag}\{\Lambda_{\mathcal R}([r_2],[r_3])\}$.
The eigenvalues of these matrices depend on the irreducible representation obtained from following tensor products:
\begin{eqnarray}
\label{tenz12}
[r_1]\otimes [r_2]&=&\oplus_{\alpha} X_{\alpha}= \oplus_{i_{12}=0}^{min(r_1,r_2)} [r_1+r_2-i_{12},i_{12}]~;~\nonumber\\
~[r_2]\otimes [r_3]&=&\oplus_{\beta} Y_{\beta}=\oplus_{i_{23}=0}^{min(r_2,r_3)} [r_2+r_3-i_{23},i_{23}]~;~\nonumber\\
~[r_1]\otimes [r_3]&=&\oplus_{\gamma} Z_{\gamma}=\oplus_{i_{13}=0}^{min(r_1,r_3)} [r_1+r_3-i_{13},i_{13}]~
\end{eqnarray}
Taking tensor product of these possible irreducible representations $X_{\alpha}, Y_{\beta}, Z_{\gamma}$ with the representation placed on third strand, we can compactly write
\begin{equation}
\label{tens}
([r_i]\otimes[r_j])\otimes [r_k]= \oplus_{\nu} {\rm dim}M^{ijk}_{\nu}~ Q_{\nu}=
\oplus_{\nu} {\rm dim}M^{ijk}_{\nu} [\ell_{\nu},m_{\nu},n_{\nu}]
\end{equation}
where $i,j,k  \in 1,2,3$ and $\ell_{\nu}+m_{\nu}+n_{\nu}= r_1+r_2+r_3$.
In order to relate the range of $i_{ij}$ in terms of $\ell_{\nu},m_{\nu}, n_{\nu}$, we highlight the relevant logical steps in the following subsection.
\subsection{Restrictions on the representations\label{s.replim}}
Let us explicitly understand the restriction of range for $i_{12}$ before we generalise for $i_{ij}$ in the tensor product (\ref{tens}). From the  relation
\begin{equation}
\label{tens1}
\oplus_{\alpha} X_{\alpha} \otimes [r_3] \equiv \oplus_{i_{12}=0}^{min (r_1,r_2)} [r_1+r_2-i_{12},i_{12}]\otimes [r_3]= \oplus_{\nu}{\rm dim}M^{123}_{\nu} [\ell_{\nu},m_{\nu},n_{\nu}]~,
\end{equation}
we can infer the following inequalities between the Young diagrams corresponding to
$X_{\alpha} \equiv[r_1+r_2-i_{12},i_{12}]$ and $Q_{\nu}\equiv[\ell_{\nu},m_{\nu},n_{\nu}]$
\begin{equation}
\begin{array}{ll}
\ell_{\nu}-(r_1+r_2-i_{12})\leq r_3, & \qquad \ell_{\nu}\geq r_1+r_2-i,
\\
m_{\nu}-i_{12}\leq r_3, & \qquad m_{\nu}\geq i_{12},
\\
r_1+r_2-i_{12}\geq m_{\nu}, & \qquad i_{12}\geq n_{\nu},
\\
n_{\nu}\leq \min(r_1,r_2,r_3), & \qquad \ell_{\nu}\geq \max(r_1,r_2,r_3)
\end{array}
\end{equation}
The above exercise for restriction of range for $i_{12}$  can be generalised for $i_{ij}$ in the tensor product(\ref{tens1}) as follows:
\begin{equation}
j_{ij,k} \leq i_{ij} \leq J_{ij,k} ~{\rm where}~ j_{ij,k}={\rm  max}(r_i,r_j-\ell_{\nu},m_{\nu}-r_k,n_{\nu}) ~;~ J_{ij,k}={\rm  min}(r_i+r_j+r_k-\ell_{\nu},r_i+r_j-m_{\nu},m_{\nu},r_i,r_j)~.
\end{equation}
Using the fact $\ell_{\nu}+m_{\nu}+n_{\nu}= r_i+r_j+r_k$, we observe
\begin{equation}
(r_i+r_j-\ell_{\nu})-(m_{\nu}-r_k)=n_{\nu}\geq 0 \Rightarrow (r_i+r_j-\ell_{\nu})\geq(m_{\nu}-r_k),
 \Rightarrow (r_i+r_j+r_k-\ell_{\nu})\geq m_{\nu}~.
\end{equation}
The above inequalities implies more stringent restriction on $j_{ij,k}$ and $J_{ij,k}$:
\begin{equation}
j_{ij,k}=\max(r_i+r_j-\ell_{\nu},n_{\nu}),\ \ \ \ J_{ij,k}=\min(r_i+r_j-m_{\nu},m_{\nu},r_i,r_j).
\end{equation}
In the following subsection, we will see that the normalised eigenvalues of
$\mathcal {R}$ are identical  whenever  the number of boxes in the Young diagrams of
$[r_1],[r_2],[r_3],Q_{\nu}$ are  reduced in a systematic way eventually modifying
three row Young diagram of $Q_{\nu}$ to two row Young diagram.

\subsection{Eigenvalues of the link $\mathcal{R}$-matrices\label{s.eigrec}}
Eigenvalues of the $\mathcal{R}$-matrix (\ref{evR1}) for $X_{\alpha} \equiv [r_1+r_2-i,i]$ is
\begin{equation}
\label{evexplicit1}
\lambda_{X_{\alpha}}=\epsilon_{X_{\alpha}}q^{r_1 r_2 +i^2 - i(r_1+r_2+1)}=(-1)^iq^{r_1 r_2 +i^2 - i(r_1+r_2+1)}.
\end{equation}
Consider an irreducible representation $Q_{\nu}$ in the decomposition (\ref{tens1}) such that
the Young diagram has non-trivial third row. That is.,  $Q_{\nu}\equiv [\ell_{\nu}, m_{\nu}, n_{\nu}>0]$. Interestingly, the multiplicity of subspace  $Q_{\nu,s}$ remains same when we
reduce the rank of  all the symmetric representations $[r_i]$'s by one as well as  change  $Q_{\nu,s} \rightarrow Q_{\nu',s}$ such that the irreducible representation
$Q_{\nu'}\equiv [\ell_\nu-1, m_{\nu}-1, n_{\nu}-1]$.

Furthermore, the eigenvalues of the $\mathcal R$ matrices under the shift of $[r_1],[r_2]\rightarrow [r_1-1],[r_2-1]$ gives new eigenvalues
\begin{equation}
\label{evexplicit2}
\lambda_{X_{\alpha^{\prime}}}=-(-1)^i q^{r_1 r_2 + i^2 - i(r_1+r_2+1)+1}.
\end{equation}
Note that ratio of the eigenvalues $\lambda_{X_{\alpha^{\prime}}}/\lambda_{X_{\alpha}}=q$
which implies that the normalised eigenvalues are indeed same. If the eigenvalue conjecture from subsection (\ref{evcj}) is correct then the Racah matrix must obey
{\small
\begin{equation}
U
\begin{bmatrix}
[r_1] & [r_2]\\
\\
[r_3] & \overline{[\ell_{\nu},m_{\nu},n_{\nu}]}
\end{bmatrix}=
U
\begin{bmatrix}
[r_1-1] & [r_2-1] \\
~&~\\
[r_3-1] & \overline{[\ell_{\nu}-1,m_{\nu}-1,n_{\nu}-1]}
\end{bmatrix}
\end{equation}
}
Following the above steps iteratively $n_{\nu}$ times, the modified $Q_{\nu'}$ Young diagram has  only two rows. The Racah matrix must remain same under such an iteration implying
{\small
\begin{equation}
\label{itera}
U
\begin{bmatrix}
[r_1] & [r_2] \\
~&~\\
[r_3] & \overline{[\ell_{\nu},m_{\nu},n_{\nu}]}
\end{bmatrix}=
U
\begin{bmatrix}
[r_1-n_{\nu}] & [r_2-n_{\nu}] \\
~&~\\
[r_3-n_{\nu}] & \overline{[\ell_{\nu}-n_{\nu},m_{\nu}-n_{\nu}]}
\end{bmatrix}
\end{equation}
}
We have obtained Racah matrices (\ref{racah3}) for some symmetric representations using highest weight method and eigenvalue hypothesis. Interestingly, these matrices (\ref{racah3})
for $Q_{\nu}=[\ell_{\nu},m_{\nu},n_{\nu}=0]$ having only two rows are agreeing with the known $U_q(sl_2)$ matrices. Hence from these results we deduce:
{\small
\begin{equation}
\label{su2racah}
U
\begin{bmatrix}
[r_1-n_{\nu}] & [r_2-n_{\nu}] \\
~&~\\
[r_3-n_{\nu}] & \overline{[\ell_{\nu}-n_{\nu},m_{\nu}-n_{\nu}]}
\end{bmatrix}=
U_{U_q(sl_2)}
\begin{bmatrix}
(r_1-n_{\nu})/2 & (r_2-n_{\nu})/2 \\
~&~\\
(r_3-n_{\nu})/2 & (\ell_{\nu}-m_{\nu})/2
\end{bmatrix}
\end{equation}
}
From eqns.(\ref{itera},\ref{su2racah}), it is clear that  the $U_q(sl_N)$ Racah matrix involving $Q_{\nu}$ (whose Young diagram has three rows) can be identified as  $U_q(sl_2)$ Racah matrix:
{\small
\begin{equation}
\label{symrac}
\boxed{
U
\begin{bmatrix}
[r_1] & [r_2] \\
~&~\\
[r_3] & \overline{[\ell_{\nu},m_{\nu},n_{\nu}]}
\end{bmatrix}=
U_{U_q(sl_2)}
\begin{bmatrix}
(r_1-n_{\nu})/2 & (r_2-n_{\nu})/2 \\
~&~\\
(r_3-n_{\nu})/2 & (\ell_{\nu}-m_{\nu})/2
\end{bmatrix}
}
\end{equation}
}
For completeness, we give the closed form expression of $U_q(sl_2)$ Racah coefficients \cite{KirResh}:

\begin{eqnarray}
\label{su2rac}
\begin{array}{r}
U_{U_q(sl_2)}\left[\begin{array}{cc|c} s_1&s_2 & i \\ s_3&s_4&
j\end{array}\right]\end{array}
 &=& \sqrt{[2i+1][2j+1]}~  (-1)^{{\sum\limits_{m=1}^4  s_m}}~
\theta(s_1,s_2,i)~
\theta(s_3,s_4,i)~\theta(s_4,s_1,j)
\theta(s_2,s_3,j)\nonumber \\
&&\times
\sum\limits_{k\geq 0}  (-1)^k[k+1]!
 \left(\frac{}{}[k-s_1-s_2-i]![k-s_3-s_4-i]![k-s_1-s_4-j]!\right.\\ &&\left.[k-s_2-s_3-j]![s_1+s_2+s_3+s_4-k]!
	[s_1+s_3+i+j-k]![s_2+s_4+i+j-k]! \frac{}{}\right)^{-1},\nonumber
\end{eqnarray}

\noindent
where the number in square bracket is called quantum number defined as  $ [n]={(q^n-q^{-n})/  (q-q^{-1})}$ and
 $$\theta(a,b,c)=\sqrt{{[a-b+c]![b-a+c]![a+b-c]!\over [a+b+c+1]!}}~. $$
Remember $s_i$'s can be integers or half-odd integers whose tensor product  will require $i,j$ to be accordingly integers or half-odd integers.

In the next section, we will explicitly work out $H_{[r_1],[r_2]}^{\L}$ for a two-component link from 3-strand braid.
\section{Colored HOMFLY polynomials for links from 3-strand braid}
\label{s.homfly}
As discussed in section \ref{s.int}, 3-strand braid group $B_3$ is generated by 2 elements $\sigma_1$ and $\sigma_2$. In order to evaluate multi-colored HOMFLY-PT of the links $H_{[r_1],[r_2],[r_3]}^{\L}$(\ref{linv})
from 3-strand braid ,  we need to compute $\R_1=\pi(\sigma_1)$ and $\R_2=\pi(\sigma_2)$.

The eigenvalues of the diagonal matrix $\Lambda_{\R_1}$ are given in eqn. (\ref{evexplicit1}). We can now explicitly write the matrix $\R_2$ by  substituting eqns.(\ref{su2rac},\ref{symrac}) into (\ref{R2}).
Hence we can evaluate $\pi(\beta_{\L})$ $\forall \beta_{\L} \in \, B_3$ colored by arbitrary symmetric representations. The braid word $\beta_{\L}$ could give one-component link (knots), two-component links
and three-component links. In fact, the two-component link contains torus knot and unknot as components.
The three-component link from $\beta_{\L}$ gives non-trivial entanglement between three unknots.
As an illustration, we compute  multi-colored  HOMFLY-PT for a two-component  link which is not just entangling of unknots in the following subsection.
\subsection{Link $L7a3$}
From the Thistlethwaite link table \cite{Thi}, we will take the simplest two-component link referred as
$L7a3$ whose non-trivial component is trefoil as drawn below:
\begin{figure}[h!]
	\centering\leavevmode
	\includegraphics[width=5.5cm]{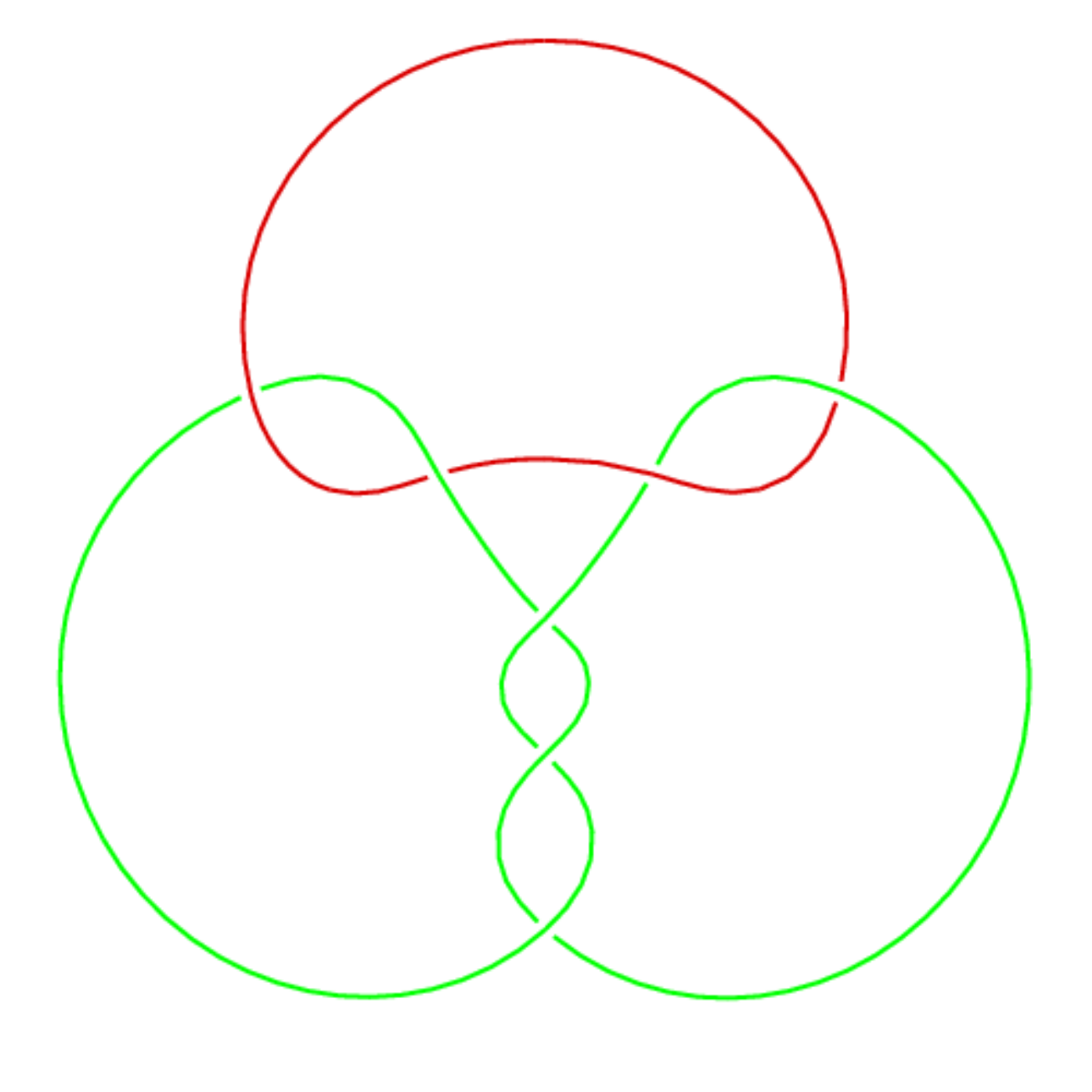}
	\caption{A picture of the link $L7a3$ from the Knotilus \cite{knotilus}}
	\label{picl7a3}
\end{figure}

The braid word $\beta_{\L} \in B_3$ for such a link is

\begin{picture}(160,80)(-70,-10)
\put(0,48){\line(1,0){18}}
\put(0,24){\line(1,0){18}}
\put(0,0){\line(1,0){34}}
\put(0,51){$[r_1]$}
\put(0,27){$[r_2]$}
\put(0,3){$[r_2]$}
\put(18,48){\line(1,-1){30}}
\put(18,24){\line(1,1){10}}
\put(34,0){\line(1,1){24}}
%
\put(42,48){\line(-1,-1){10}}
\put(64,0){\line(-1,1){12}}
\put(64,24){\line(1,0){7}}
%
\put(42,48){\line(1,0){30}}
\put(64,0){\line(1,0){52}}
\put(57,24){\line(1,0){16}}
\put(72,48){\line(1,-1){24}}
\put(72,24){\line(1,1){10}}
\put(96,48){\line(-1,-1){10}}
%
\put(42,48){\line(1,0){30}}
\put(96,48){\line(1,0){10}}
\put(96,24){\line(1,0){10}}
\put(106,24){\line(1,1){10}}
\put(106,48){\line(1,-1){24}}
%
\put(130,48){\line(-1,-1){10}}
\put(130,48){\line(1,0){10}}
\put(130,24){\line(1,0){10}}
%
\put(140,48){\line(1,-1){28}}
\put(140,24){\line(1,1){10}}
\put(164,48){\line(-1,-1){10}}
\put(164,48){\line(1,0){10}}
\put(174,48){\line(1,-1){24}}
\put(198,24){\line(1,0){30}}
\put(64,0){\line(1,0){90}}
\put(154,0){\line(1,1){32}}
\put(188,0){\line(-1,1){15}}
\put(188,0){\line(1,0){41}}
\put(201,48){\line(-1,-1){12}}
\put(201,48){\line(1,0){28}}
%
\end{picture}

There are two diagonal $\mathcal{R}$-matrices:
$\Lambda_{\mathcal{R}}([r_2],[r_2])$, $\Lambda_{\mathcal{R}_2}([r_2],[r_2])$.
$\Lambda_{\mathcal{R}}([r_1],[r_2])$ stands for crossings of representations $[r_1]$ and $[r_2]$  and $\Lambda_{\mathcal{R}}([r_2],[r_2])$ that stands for crossings of representations $[r_2]$ and $[r_2]$. The two Racah matrices
\be
U_{1} := U\left[\begin{array}{cc} [r_2]&~[r_2] \\ ~[r_1]&\bar Q \end{array}\right] \quad \text{and} \quad U_{2}:=U\left[\begin{array}{cc} [r_2]~&~[r_1] \\ ~[r_2]&\bar Q \end{array}\right]
\ee
correspond accordingly to the placements of representations $[r_2],[r_2],[r_1]$ and $[r_2],[r_1],[r_2]$. The answer for the HOMFLY polynomial (\ref{linv}) is then given by:
\begin{equation}
\label{hmfl7a3}
H_{[r_1],[r_2]}^{\text{L7a3}}  =
\sum\limits_{Q} \text{tr}\left(\Lambda_{\mathcal{R}}([r_1],[r_2])\cdot U_{2}^{\dagger}\cdot \Lambda^{-1}_{\mathcal{R}}([r_1],[r_2])\cdot U_{1}\cdot \Lambda_{\mathcal{R}}([r_2],[r_2])^{3}\cdot U_{1}^{\dagger}\cdot \Lambda^{-1}_{\mathcal{R}}([r_2],[r_1])\cdot U_{2}\cdot \Lambda_{\mathcal{R}}([r_1],[r_2]) \right) \cdot s_Q^*(A,q),
\end{equation}
where $Q\in [r_1]\otimes [r_2]\otimes [r_2]$. Note that the second component colored by representation $[r_2]$  is trefoil.

We have performed calculations for some values of $[r_1]$ and $[r_2]$ using the eigenvalues of $\mathcal R$ as well as Racah matrix elements discussed in section \ref{s.3br}.
The multi-colored HOMFLY-PT (\ref{hmfl7a3}) for arbitary colors $[r_1]$ and $[r_2]$ seems to have a neat form involving quantum number factorial $[n]!=[n][n-1]\ldots [1]$ as well as numbers in parenthesis defined as
\begin{equation}
\{x\}=x-{1\over x}~~;~~
D_k={\{Aq^k\}\over\{q\}}~.
\end{equation}
The explicit expression for this two-component link is
\begin{equation}\label{g2}
\dfrac{H_{[r_1],[r_2]}^{\text{L7a3}}}{s_{[r_1]}^*\cdot s_{[r_2]}^*}=T_{[r_2]}(q,A)+\sum_{k=1}^{min(r_1,r_2)} {[r_1]![r_2]!\over [r_1-k]![r_2-k]!} {\{q\}^{3k}\over A^{3r_2}}\frac{D_{-1}}{D_{r_2-1}}
{\prod_{n=1}^k D_{r_1+n-1} \prod_{m=0}^{r_2-k-1}D_{2k+m}\over \prod_{i=0}^{r_2-k-1}D_{k+i-1}} \cdot G_{k,r_2}(q,A)~,
\end{equation}
where $T_{[r_2]}(q,A)$) refers to  (reduced) colored HOMFLY polynomial of the trefoil
in the topological framing colored with $[r_2]$ and $G_{k,r_2}$ can be derived using
\begin{equation}
G_{1,r_2}(q,A)=\sum_{i=1}^{r_2} Q_{i,r_2}(q,A) ~{\rm where}~
Q_{i,r_2}(q,A)={[2(i+1)]\over [i+1]}{[r_2-1]!\over [i-1]![r_2-i]!}{\{q\}^{i-1}\over A^{i}q^{2r_2^2-i^2/2-5i/2+1}} \prod_{j=2}^iD_{r_2+j-1}
\end{equation}
as follows:
\begin{equation}
\label{G1s}
G_{k,r_2}(q,A)=\sum_{i=1}^{r_2-k+1}\left(\prod_{n=1}^{k-1}{[r_2-i-n+1]\over [r_2-n]}\right)q^{(1-k)(k+2i-2)/2} \left[G_{1,r_2}(q,A)\right]_{1-2i}^AA^{r_2-k+2-2i}
\end{equation}
where $[G]_p^A$ denotes the coefficient of the $p$-th degree of the (Laurent) polynomial $G$ of $A$. Equivalently, the form of $G_{k,r_2}(q,A)$ can be written as double sum :
\begin{eqnarray}
G_{k,r_2}(q,A)&=&\sum\limits_{i=1}^{r_2} P_{i,k,r_2}(q,A) ~~{\rm where}~~\\
P_{i,k,r_2}(q,A)&=&{q^{i(i+5)/2}\over q^{2r_2^2+{k(k-1)\over 2}+1}A^{2i+k-r_2-2}}{[2i{+}2]\over [i+1]}{[r_2-1]!\over [r_2{-}i]![i{-}1]!}\prod_{p=r_2+k}^{r_2+i-1} \hspace{-1mm} \tilde D_p
\sum_{n=1}^{\text{min}(i,k)} \prod_{m=r_2+n}^{r_2{+}\text{min}(i,k){-}1} \hspace{-2mm} \tilde D_m\nonumber\\
~&~&\times \prod_{j=1}^{n-1}{[i{-}j][k{-}j]\over [j][r_2-j]}{1\over q^{j+k+i-2}}~.\nonumber
\end{eqnarray}
Here $\tilde D_m=A\{q\}D_m = A^2q^{m}-q^{-m}$. Remember $[0]!=1$ and terms inside $\prod_{j=a}^b$
involving $b<a$ are set to one.

Another way to evaluate the HOMFLY polynomial for this link is described in \cite{MMMtang}.

\section{Conclusion}
In this paper, we study invariants of  links from 3-strand braids where the strands are colored by arbitrary symmetric representations. We have presented the construction of
multi-colored invariants $H_{[r_1],[r_2],[r_3]}^{\L}$ (\ref{HMF1}) for these links,where the component knots are colored by symmetric $[r_1],[r_2],[r_3]$ representations, using the quantum $\mathcal R$ as well as $U_q(sl_N)$ Racah matrices. In the literature, the matrix elements of these Racah matrices are not available for arbitrary representations. So we cannot explictly write the multi-colored  HOMFLY-PT in terms of variables $q,A$. Hence the main theme of this paper was to determine these Racah matrix elements.

We have given a formal proof, using eigenvalue hypothesis, to identify the Racah matrices with quantum $U_q(sl_2)$ Racah matrices. As the $U_q(sl_2)$ Racah matrix coefficients are known \cite{KirResh}, we could evaluate the multi-colored HOMFLY-PT for two and three-component links colored by different symmetric representations. For concreteness, we have presented the results for a two-component link $L7a3$ (\ref{hmfl7a3})-(\ref{g2}).

The extension of the proof for  $m>3$-strand braids looks plausible. It appears that these Racah matrices may be identified with $U_q(sl_{m-1})$. We hope to get insight on Racah matrices obtainable from highest weight approach \cite{multeg} for some low rank  symmetric representations of $U_q(sl_N)$ placed on component knots.  The results could lead to writing closed form expession for $U_q(sl_{N\geq 3})$ Racah coefficients for arbitrary symmetric representations. Extension to multiplicity free rectangular representations must also be addressed for $3$-strand and $m>3$-strand braids.  We will pursue these issues in  future so that multi-colored HOMFLY-PT for links from $m$-strand braids become computable.

\section*{Acknowledgements}
Our work was partly supported by the grant of the Foundation for the Advancement of Theoretical Physics ``BASIS" (A.Mor. and A.S.),  by grant 16-31-60082-mol-a-dk (A.S.), by RFBR grants 16-01-00291 (A.Mir.), 16-02-01021 (A.Mor.) and 17-01-00585 (An.Mor.), by joint grants 17-51-50051-YaF, 18-51-05015-Arm-a (A.M's, A.S.), 18-51-45010-Ind-a (A.M's, A.S.). PR, VKS and SD acknowledge DST-RFBR grant(INT/RUS/RFBR/P-231) for support. SD would like to thank CSIR for research fellowship.

\end{document}